\documentclass[twocolumn]{el-author}
\newcommand{\figref}{Fig.~\ref}

\date{\today}

\usepackage{cite,graphicx,amsmath,psfrag,bm}
\usepackage{subfigure}
\usepackage[usenames,dvipsnames]{xcolor}
\usepackage{mathabx}
\usepackage{cancel}
\usepackage{epstopdf}
\usepackage{pstool}
\usepackage{caption}
\begin{document}
\title{Planar Reflective Phaser and Synthesis for Radio Analog Signal Processing (R-ASP)}

\author{Lianfeng Zou, Qingfeng Zhang and Christophe Caloz}

\abstract{A planar reflective phaser based on an open-ended edge-coupled-line structure is proposed. This phaser is the first reported phaser that combines the benefits of high resolution, inherent to cross-coupled resonator reflective phasers, and of compactness, inherent to planar circuits. A \mbox{4-ns swing 4.9-5.5 GHz quadratic phase (linear group delay) $4^\text{th}$-order} microstrip phaser is synthesized and experimentally demonstrated. Given its advantages, this phaser may find vast applications in Radio Analog Signal Processing (R-ASP) systems.
\vspace{-0.1in}}
\maketitle

\section{Introduction}

Emerging Radio Analog Signal Processing (R-ASP) technology, which consists in manipulating signals in their pristine analog and real-time form, represents a novel and promising approach for future millimeter-wave and terahertz communication, radar, sensing and imaging systems~\cite{Jour:2013_MwMag_Caloz}. Compared to conventional digital signal processing (DSP) based technologies, R-ASP offers indeed the benefits of higher speed, lower cost and lower consumption, especially when deployed at millimeter-wave frequencies. R-ASP applications reported so far include pulse position modulators~\cite{Nguyen_MWCL_08_2008}, compressive receivers~\cite{JOUR:2009_TMTT_Abielmona}, real-time spectrum analyzers~\cite{JOUR:2003_TMTT_Laso}, real-time spectrogram analyzers~\cite{JOUR:2009_TMTT_Gupta}, waveform synthesisers~\cite{JOUR:1991_TMTT_Leonard}, and chipless RFIDs~\cite{JOUR:2011_AWPL_Gupta}, to name just a few.

The core element in R-ASP is the ``phaser''~\cite{Jour:2013_MwMag_Caloz}, a component providing a highly flexible and controllable frequency dependent group delay response, and which may be transmission-type or reflection-type structures. In the former catergory is the C-section phaser~\cite{JOUR:2010_TMTT_Gupta}, whose processing resolution, defined as the group delay swing -- bandwidth product, \mbox{$\sigma=\Delta\tau\Delta\omega$}, is restricted by practically achievable coupling levels. A CRLH C-section, leveraging CRLH coupling enhancement, was proposed was proposed in~\cite{JOUR:2012_TMTT_Shulabh} to boost this resolution, but this was at the expense of higher design complexity. On the other hand, reflection-type phasers provide higher resolution~\cite{JOUR:2012_TMTT_Zhang}, but must be combined with circulators or a couplers for reflection to transmission response transformation. They may be realized in planar Bragg grating technology~\cite{JOUR:2003_TMTT_Laso}, but this approach suffers from group delay oscillations due to multiple internal reflections. To remedy this issue, a coupled-resonator based reflective phaser was introduced in~\cite{JOUR:2012_TMTT_Zhang}. This phaser exhibits excellent characteristics but it was realized in bulky rectangular waveguide technology.

This paper presents a planar alternative to the reflective phaser of~\cite{JOUR:2012_TMTT_Zhang}. The proposed phaser is an open-ended edge-coupled transmission line structure, which benefits from all the benefits of planar circuits.

\vspace{-0.1in}

\section{Principle}

We select here a side-coupled half-wavelength resonator configuration, where each of the cascaded resonators are juxtaposed with the two resonators, as this configuration is more compact than the end-coupled resonator one and does not suffer from even-harmonic parasitic response~\cite{BK:1980_Matthaei}. Moreover, we choose an edge coupling, as opposed to broadside coupling, for circuit uniplanarity~\cite{BK:2011_Pozar,BK:1980_Matthaei}.

\begin{figure}[hbt!]
    \centering
    \psfragfig*[width=1\linewidth,trim={-0.75in -0.25in -0.75in -0.25in}]{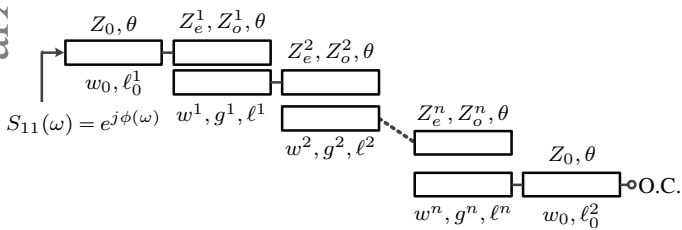}{
   \psfrag{s}[tc][bl][1]{$S_{11}(\omega)=e^{j\phi(\omega)}$}
   \psfrag{g}[bc][bc][1]{$Z_0, \theta$}
   \psfrag{h}[bc][bc][1]{$Z_e^1, Z_o^1, \theta$}
   \psfrag{i}[bc][bc][1]{$Z_e^2, Z_o^2, \theta$}
   \psfrag{j}[bc][bc][1]{$Z_e^n, Z_o^n, \theta$}
   \psfrag{k}[bc][bc][1]{$Z_0, \theta$}
    \psfrag{f}[tc][tc][1]{$w_0, \ell_0^1$}
   \psfrag{a}[tc][tc][1]{$w^1, g^1, \ell^1$}
   \psfrag{b}[tc][tc][1]{$w^2, g^2, \ell^2$}
   \psfrag{c}[tc][tc][1]{$w^n, g^n, \ell^n$}
   \psfrag{d}[tc][tc][1]{$w_0, \ell_0^2$}
   \psfrag{o}[l][l][1]{O.C.}
   }
    \caption{Topology of an $n^{\text{th}}$-order reflective planar phaser composed of edge-coupled resonators.\vspace{-0.1in}}
    \label{FIG:Topology}
\end{figure}

Figure~\ref{FIG:Topology} shows the topology of an $n^{\text{th}}$-order open-ended reflective phaser based on edge-coupled resonators. Each coupled-section is a two-port network obtained by opening two diagonally opposite ports of a four-port coupled-line coupler ~\cite{BK:2011_Pozar}. The key electrical parameters for the $i^{\text{th}}$ coupled-section are the even- and odd-mode characteristic impedances, $Z_e^i$ and $Z_o^i$. The first and last sections have the same characteristic impedance as that of the external system, $Z_0$, and all the sections have an electrical length of $\theta=\pi/2$ at the resonance frequency, $\omega_0$.

The synthesis of the desired group delay response, \mbox{$\tau(\omega)=-\partial\phi(\omega)/\partial\omega$} for \mbox{$\omega_L\leqslant\omega\leqslant\omega_H$}, where $\phi(\omega)=\angle\{S_{11}(\omega)\}$ and where $|S_{11}(\omega)|=1$ due to the purely reflective nature of the device, follows the procedure established in~\cite{JOUR:2012_TMTT_Zhang}, based on the equivalent circuits shown in~\figref{FIG:Topology}~\cite{BK:2011_Pozar}.
\begin{figure}[hbt!]
    \centering
    \subfigure[]{
        \label{FIG:EqCct:BP}
        \psfragfig*[width=1\linewidth]{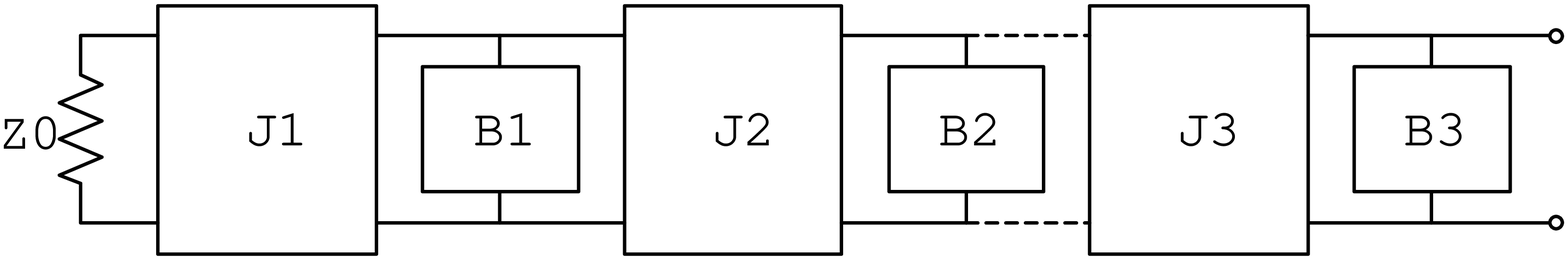}{
        \psfrag{Z0}[r][r][0.85]{$Y_0$}
        \psfrag{J1}[c][c][0.85]{$J_{0,1}(\omega)$}
        \psfrag{J2}[c][c][0.85]{$J_{1,2}(\omega)$}
        \psfrag{J3}[c][c][0.85]{$J_{n-1,n}(\omega)$}
        \psfrag{B1}[c][c][0.85]{$B_1(\omega)$}
        \psfrag{B2}[c][c][0.85]{$B_2(\omega)$}
        \psfrag{B3}[c][c][0.85]{$B_n(\omega)$}
        }
    }
    \subfigure[]{
        \label{FIG:EqCct:LP}
        \psfragfig*[width=1\linewidth]{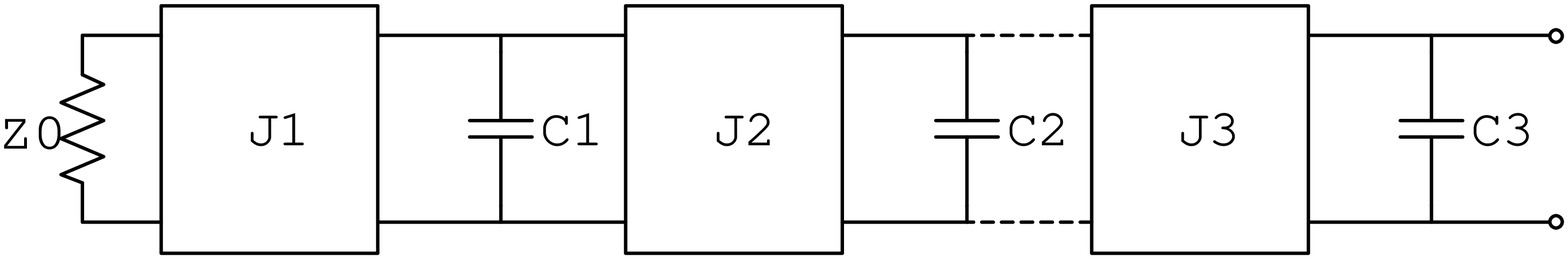}{
        \psfrag{Z0}[r][r][0.85]{$Y_0$}
        \psfrag{J1}[c][c][0.85]{$J_{0,1}(\Omega)$}
        \psfrag{J2}[c][c][0.85]{$J_{1,2}(\Omega)$}
        \psfrag{J3}[c][c][0.85]{$J_{n-1,n}(\Omega)$}
        \psfrag{C1}[c][c][0.85]{$C_1$}
        \psfrag{C2}[c][c][0.85]{$C_2$}
        \psfrag{C3}[c][c][0.85]{$C_n$}
        }
    }
   \subfigure[]{
        \label{FIG:EqCct:LLN}
        \psfragfig*[width=1\linewidth]{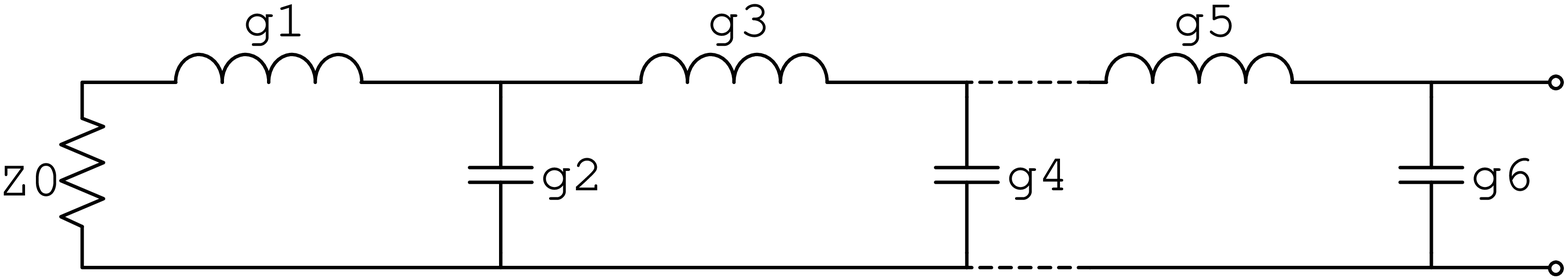}{
        \psfrag{Z0}[r][r][1]{$g_0$}
        \psfrag{g1}[c][c][1]{$g_1$}
        \psfrag{g2}[c][c][1]{$g_2$}
        \psfrag{g3}[c][c][1]{$g_3$}
        \psfrag{g4}[c][c][1]{$g_4$}
        \psfrag{g5}[c][c][1]{$g_{n-1}$}
        \psfrag{g6}[c][c][1]{$g_n$}
        }
    }
    \caption{Equivalent circuits for the planar reflective phaser of~\figref{FIG:Topology}. (a)~Bandpass distributed equivalent circuit, with response $S_{11}(\omega)=\exp[j\phi(\omega)]$. (b)~Lowpass (\mbox{$0\leqslant\Omega\leqslant1$}) equivalent circuit with admittance inverters, $J_{i-1,i}$, \mbox{$i=1,2,\ldots,n$}. (c)~Corresponding fully-lumped prototype.}\vspace{-0.1in}
    \label{FIG:EqCct}
\end{figure}

Figure~\ref{FIG:EqCct:BP} is the bandpass equivalent circuit of the phaser, with center frequency $\omega_0$, where the half-wavelength resonators are modeled by distributed susceptances, $B_i(\omega)$, and the couplings between these resonators are modeled by admittance-inverters, $J_{i,i+1}(\omega)$. Figure~\ref{FIG:EqCct:LP} is the normalized frequency, \mbox{$0\leqslant\Omega\leqslant1$}, low-pass equivalent circuit, with admittance-inverters and lumped capacitors, and \figref{FIG:EqCct:LLN} is the corresponding prototype, with unity input impedance \mbox{$g_0=1$}.

The synthesis is performed as follows. First, the bandpass frequencies and corresponding phases, pertaining to \figref{FIG:EqCct:BP}, are discretized, $\omega\rightarrow\omega_i$ and $\phi\rightarrow\phi_i$, with \mbox{$i=0,1,2,\ldots,n$}. Second, $\omega_i$ and $\phi_i$ are mapped onto their lowpass counterparts, $\Omega_i$ and $\Phi_i$ with \mbox{$0\leqslant\Omega_i\leqslant1$} corresponding to the \figref{FIG:EqCct:LP}, using the mapping function \mbox{$\Omega_i=\tan(\pi\omega_i/\omega_0)/\tan(\pi\omega_n/\omega_n)$}. Third, a Hurwitz polynomial, $H(s)$ where $s=j\Omega$, realizing the phase function $\Phi(\Omega)$ at the discretized points is constructed using a recurrence formula (given in~\cite{JOUR:2012_TMTT_Zhang}). Since $|S_{11}|=1$ and $\angle\{S_{11}\}=\phi(\omega)$, $S_{11}$ has the polynomial form $S_{11}(s)=H(-s)/H(s)$, so that $z(s)=(1-S_{11})/(1+S_{11})=[H(s)-H(-s)]/[H(s)+H(-s)]=H_o(s)/H_e(s)$, where $H_o(s)$ and $H_e(s)$ are the even and odd parts of $H(s)$, respectively. Then, fourth, this polynomial expression of $z(s)$ is mapped onto the input impedance of~\figref{FIG:EqCct:LLN}, which provides the parameters $g_i$. Fifth, the parameters in~\ref{FIG:EqCct:LP} are computed from the $g_i$'s as

\begin{subequations}\label{EQ:BP_Elements}
    \begin{equation}\label{EQ:BP_Elements:J01}
      J_{0,1}=\sqrt{\dfrac{Y_0C_1}{g_0g_1}}=Y_0\sqrt{\dfrac{B}{g_1}},\quad
      J_{i-1,i}=\sqrt{\dfrac{C_{i-1}C_{i}}{g_{i-1}g_{i}}}=Y_0\dfrac{B}{\sqrt{g_{i-1}g_{i}}},
    \end{equation}
    with
    \begin{equation}\label{EQ:BP_Elements:B}
      C_i =Y_0\tan\left(\pi\dfrac{\omega_n}{\omega_0}\right),\quad
      B_i=B = \tan\left(\pi\dfrac{\omega_n}{\omega_0}\right),\quad\forall i,
    \end{equation}
\end{subequations}
where $Y_0=1/Z_0=50\text{ $\Omega$}$. The parameters $C_i$ and $B_i$ are the same for all $i's$ because all the resonators are synchronously tuned at $\omega_0$, which may be chosen to be either $\omega_L$ or $\omega_H$. Finally, sixth, the electrical parameters $Z_e^i$ and $Z_o^i$ are calculated by inserting~\eqref{EQ:BP_Elements:J01} into the following equations~\cite{BK:2011_Pozar}:
\begin{subequations}\label{EQ:IMPL}
    \begin{equation}\label{EQ:IMPL_Ze}
      Z_e^i = \frac{1}{Y_0}\left[1+\frac{J_{i-1,i}}{Y_0}+\left(\frac{J_{i-1,i}}{Y_0}\right)^2\right],
    \end{equation}
    \begin{equation}\label{EQ:IMPL_Zo}
      Z_o^i = \frac{1}{Y_0}\left[1-\frac{J_{i-1,i}}{Y_0}+\left(\frac{J_{i-1,i}}{Y_0}\right)^2\right],
    \end{equation}
\end{subequations}
from which the layout parameters $(w^i,g^i,\ell^i)$ in Fig.~\ref{FIG:Topology} may be determined by conventional microwave techniques~\cite{BK:2007_Mongia}.

\section{Numerical and Experimental Demonstration}

\begin{figure}[hbt]
    \centering
    \includegraphics[width=1\linewidth]{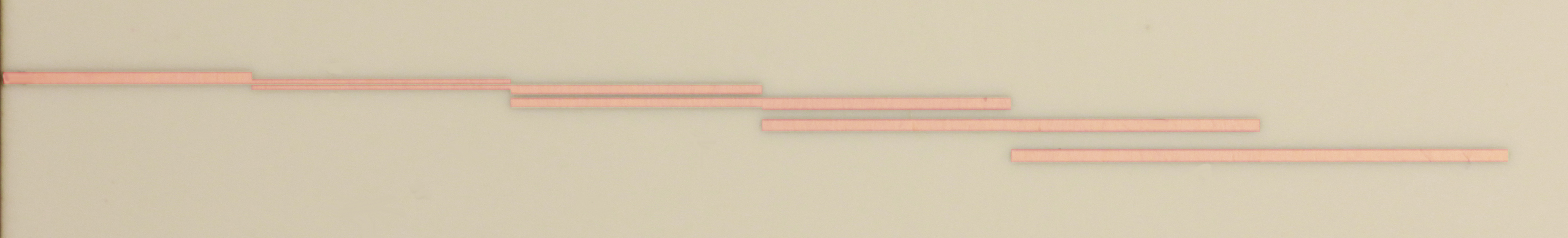}
    \caption{Layout of the prototype, having the dimensions listed in Tab.~\ref{TAB:PhyDim}.\vspace{-0.1in}}
    \label{FIG:Layout}
\end{figure}

To validate the concepts of the previous section, a microstrip implementation of the proposed planar reflective phaser is presented here. The resonators are etched on an $\text{Al}_2\text{O}_3$ substrate with relative permittivity \mbox{$\varepsilon_r = 9.9$}, loss-tangent $\tan\delta = 0.001$, substrate thickness $10\text{ mil}$ and metal thickness $1\text{ $\mu$m}$. The prescribed phaser response is a \mbox{4 ns-swing} up-chirp linear group delay over the range \mbox{$4.9-5.5\text{ GHz}$}. Accordingly a~\mbox{$4^\text{th}$-order} phaser was synthesized. The synthesized geometrical parameters, indicated in~\figref{FIG:Topology}, are listed in Tab.~\ref{TAB:PhyDim}. The physical lengths the coupled-line section are slightly smaller than the effective quarter-wavelength at $\omega_0$ after taking into account the open-end capacitive effect~\cite{BK:2011_Hong}. The first and last sections are $50\text{ $\Omega$}$ transmission lines, whose dimensions are $w_0=9.543268\text{ mil}$, $\ell_0^1=208.201969\text{ mil}$ and $\ell_0^2=205.260859\text{ mil}$. A photograph o the layout is shown in \figref{FIG:Layout}.
\begin{table}[h!t]
    \setlength{\belowcaptionskip}{7pt}
     \setlength\tabcolsep{6pt}
    \caption{Dimensions (unit: mil) for the \mbox{$4.9-5.5\text{ GHz}$} $4^{\text{th}}$ order\\  microstrip  reflective phaser with $4 \text{ ns}$ group delay swing.}
    \label{TAB:PhyDim}
    \centering
    \begin{tabular}{l|cccccc}
        & I & II & III & IV \\[1pt]
      \hline\\[-8pt]
       $w^i$ & 3.115516 & 7.303504 & 8.898937 & 9.300669\\[1pt]
       $g^i$ & 1.459563 & 3.802933 & 9.132362 & 16.092402\\[1pt]
       $\ell^i$ & 213.085436 & 207.405863 & 205.772824 & 205.371623
    \end{tabular}
    \vspace{-0.2in}
 \end{table}

\begin{figure}[hbt]
\setlength{\abovecaptionskip}{0pt}
\setlength{\belowcaptionskip}{-20pt}
    \centering
    \subfigure[]{
        \label{FIG:Result_Gd}
        \psfragfig*[width=0.9\linewidth]{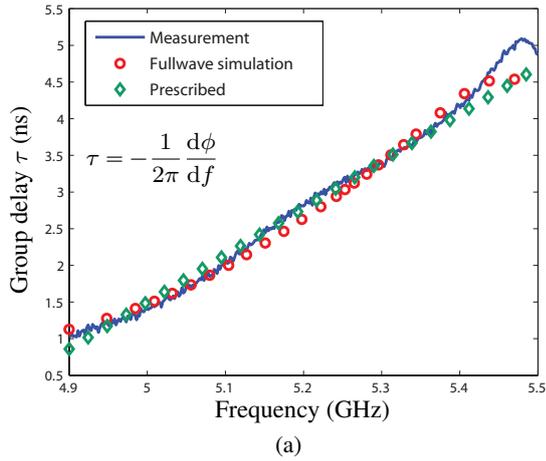}{
        \psfrag{G}[Bc][Bc][1.1]{Group delay $\tau$ (ns)}
        \psfrag{e}[l][l][1.1]{$\tau=-\dfrac{1}{2\pi}\dfrac{\mathrm{d}\phi}{\mathrm{d}f}$}
        \psfrag{F}[tc][tc][1.1]{Frequency (GHz)}
        }
    }
    \subfigure[]{
        \label{FIG:Result_Mag}
        \psfragfig*[width=0.9\linewidth]{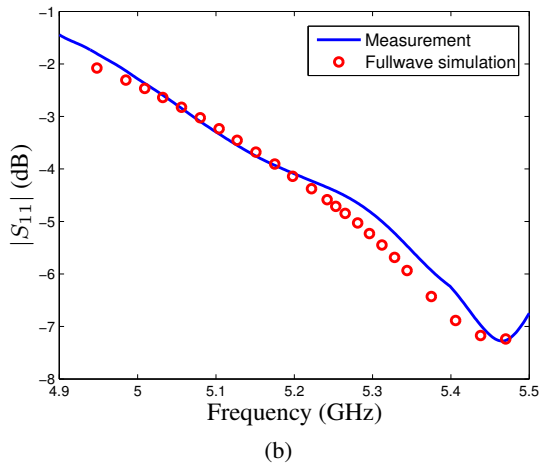}{
        \psfrag{S}[Bc][Bc][1.1]{$|S_{11}|$ (dB)}
         \psfrag{F}[tc][tc][1.1]{Frequency (GHz)}
         }
    }
    \caption{Measurement results compared against fullwave simulation and prescribed (only shown for group delay response) ones, (a) group delay response, (b) magnitude response $|S_{11}|$.\vspace{0.1in}}
    \label{FIG:Result}
\end{figure}

The measured group delay response and reflection coefficient are plotted in Figs.~\ref{FIG:Result_Gd} and \figref{FIG:Result_Mag}, respectively. The prescribed group delay is also shown for comparison in the former case. Good agreement between measurement and simulation results is observed.

\section{Conclusion}

A reflective phaser based on an open-ended edge-coupled line structure has been demonstrated. This phaser combines the benefits of high resolution inherent to cross-coupled resonator reflective phasers and of compactness of planar circuits. It may therefore find vast applications in R-ASP.

\nocite{*}\bibliography{IEEEabrv,reference2}
\end{document}